\documentclass[aps,showpacs, twocolumn,twoside]{revtex4}
\usepackage{epsfig,amsmath,amssymb}
\expandafter\ifx\csname package@font\endcsname\relax\else
 \expandafter\expandafter
 \expandafter\usepackage
 \expandafter\expandafter
 \expandafter{\csname package@font\endcsname}%
\fi

\def\ket#1{|#1\rangle}
\def\bra#1{\langle#1|}

\def\bz{\beta_0}

\begin{document}

\preprint{APS/123-QED}

\title{Regularity and chaos at critical points of first-order 
quantum phase transitions}
\author{M.~Macek}
\author{A.~Leviatan}

\affiliation{Racah Institute of Physics, The Hebrew University, 
Jerusalem 91904, Israel}
\date{\today}

\begin{abstract}
We study the interplay between regular and chaotic dynamics at 
the critical point of a generic first order quantum phase transition 
in an interacting boson model of nuclei. 
A classical analysis reveals a distinct behavior 
of the coexisting phases in a broad energy range. 
The dynamics is completely regular in the deformed phase and, 
simultaneously, strongly chaotic in the spherical phase. 
A quantum analysis of the spectra 
separates the regular states from the irregular ones, 
assigns them to particular phases 
and discloses persisting regular rotational bands in the deformed region. 
\end{abstract}


\maketitle

Quantum phase transitions (QPTs) at zero temperature  
are structural changes of the system's ground state 
resulting from a variation of 
parameters $\lambda$ in the quantum Hamiltonian 
$\hat{H}(\lambda)$~\cite{ref:Gilm79}. Aspects of QPTs and their 
finite-$N$ precursors are currently 
attracting considerable theoretical and experimental 
interest, as they occur in diverse dynamical systems 
including spin lattices~\cite{ref:Sachdev99}, 
ensembles of ultracold atoms~\cite{ref:Grei02} and  
atomic nuclei~\cite{ref:Cejn10}. 

The particular type of QPT is reflected in the topology of the 
underlying mean-field (Landau) potential $V(\lambda)$. 
In first order (discontinuous) QPTs, $V(\lambda)$ develops 
multiple minima, corresponding to distinct phases 
(characterized by distinct values of an order parameter) 
that coexist in a range of $\lambda$-values 
and become degenerate at the critical point $\lambda=\lambda_c$. 
In higher order (continuous) QPTs, the single minimum of 
$V(\lambda)$ shifts continuously but non-analytically 
to a different value of the order parameter 
at $\lambda=\lambda_c$, without phase coexistence. 

Interacting many-body systems undergoing QPTs provide a fertile 
ground for studying the emergence of quantum chaos. A typical 
example, is the recent analysis in quantum optics models of $N$ two-level 
atoms interacting with a 
single-mode radiation field~\cite{ref:Emar03}, where 
the onset of chaos is triggered by continuous QPTs. 
In this communication, we turn our attention to systems with discontinuous 
QPTs, where multiple-well potentials and phase coexistence provide even 
a richer environment to study the interplay of order and chaos. Indeed, 
studies of 2D single-particle models indicate that 
the classical motion can be dissimilar (regular or chaotic) 
in different local minima~\cite{ref:Bere04}. 
In the present work, we show that such mixed form of dynamics and 
non-uniform onset of chaos arises from rotational-invariant 
Hamiltonians in a system of $N$ interacting constituents, undergoing a 
first-order QPT. A marked separation between regular and chaotic 
dynamics is observed in both classical and quantum analyses. 

As a concrete example, we adopt the interacting boson model 
(IBM)~\cite{ref:Iach87}, widely used 
in the description of quadrupole collective states in 
nuclei, in terms of a system of $N$ monopole ($s$) and
quadrupole ($d$) bosons, representing valence nucleon pairs. 
The total boson number $N$ and angular momentum $L$ are conserved, 
which enables to solve the model exactly by numerical diagonalization of 
finite matrices for any fixed $N$ and $L$. 
The model accommodates both continuous and discontinuous quantum shape-phase 
transitions~\cite{ref:Diep80}, which are manifested empirically between 
spherical and deformed nuclei~\cite{ref:Cejn10}. 

Apart from kinetic rotational terms, the most general two-body 
IBM Hamiltonian at the 
critical point of 
a first order QPT, can be transcribed in the form~\cite{ref:Lev06}
\begin{eqnarray}
\label{eq:Hcri}
\hat{H}_\mathrm{cri} = 
h_2 P^\dag_2(\beta_0)\cdot\tilde{P}_2(\beta_0)/N(N-1) ~,
\end{eqnarray}
where 
$P^{\dagger}_{2\mu}(\beta_0) = 
\beta_{0}\,s^{\dagger}d^{\dagger}_{\mu} + 
\sqrt{7/2}\,\left( d^{\dagger} d^{\dagger}\right )^{(2)}_{\mu}$, 
$\tilde{P}_{2\mu}(\beta_0)=(-1)^{\mu}P_{2,-\mu}(\beta_0)$ 
and the dot implies a scalar product. 
To facilitate the comparison with the classical limit, involving large $N$, 
the critical-point Hamiltonian~(\ref{eq:Hcri}) is scaled by $N(N-1)$. 
For $\bz>0$, $\hat{H}_\mathrm{cri}$ annihilates 
two distinct degenerate ground states, which have the form of 
static condensates $\ket{\beta;N}=(N!)^{-1/2}\Gamma_c^\dag(\beta)^N\ket{0}$ 
of $N$ bosons,  
$\Gamma_c^\dag(\beta)=[\beta d_0^\dag + \sqrt{2-\beta^2}s^\dag]/\sqrt{2}$, 
with (i)~$\beta = 0$ (spherical $s$-boson condensate) and 
(ii)~$\beta=\sqrt{2}\bz(1+\bz^2)^{-1/2}$ (deformed condensate) and 
correspond to the two coexisting shape phases of the 
nucleus. The parameter $\beta_0$ in  $\hat{H}_\mathrm{cri}$ 
determines the equilibrium deformation 
in the deformed phase as well as the height of the potential barrier 
between the two phases (see below). 

The classical limit of the IBM is obtained through the use of coherent 
states~\cite{ref:Hatc82}. This amounts to 
replacing $(s^{\dagger},\,d^{\dagger}_{\mu})$ by six 
c-numbers $(\alpha_{s}^{*},\,\alpha_{\mu}^{*})$ rescaled 
by $\sqrt{N}$ and taking $N\rightarrow\infty$, with $1/N$ playing the 
role of $\hbar$. Number conservation ensures that 
phase space is 10-dimensional and can be phrased in terms of 
two shape (deformation) variables, three orientation (Euler) angles 
and their conjugate momenta. Chaotic properties of the IBM have been 
studied extensively~\cite{ref:AW93}, albeit, with a simplified Hamiltonian, 
giving rise to an extremely low, hence non-generic, barrier. 
The Hamiltonian of Eq.~(\ref{eq:Hcri}) can accommodate 
a barrier with an adjustable height, hence allows a systematic study of 
its impact on the dynamics. 
In the classical analysis presented below we consider, for simplicity, the 
dynamics of $L=0$ vibrations, for which only two 
degrees of freedom are active. 
The rotational dynamics with $L>0$ is examined in 
the subsequent quantum analysis. 

For the particular case of the 
Hamiltonian~(\ref{eq:Hcri}), constrained to $L=0$, the 
above procedure yields the following classical Hamiltonian 
\begin{eqnarray}\label{eq:Hcl}
&&H_\mathrm{cri}/h_2 = 
{\cal H}_{d,0}^2 + p_\gamma^2 + \bz^2(1-{\cal H}_{d,0}){\cal H}_{d,0}
\nonumber \\
&&\quad+ \bz\sqrt{(1-{\cal H}_{d,0})/2}  
\nonumber\\
&&\quad\;\; \times 
[(p_\gamma^2/\beta - \beta p^2_{\beta} - \beta^3) \cos{3\gamma} 
+ 2p_{\beta} p_\gamma \sin{3\gamma}]~.\quad\;
\end{eqnarray}
Here the coordinates $\beta\in[0,\sqrt{2}]$, $\gamma\in[0,2\pi)$ and their 
canonically conjugate momenta $p_\beta\in[0,\sqrt{2}]$ and $p_\gamma\in[0,1]$ 
span a compact classical phase space. The term 
${\cal H}_{d,0}\equiv(T+\beta^2)/2$, 
with $T=p_{\beta}^2+p_\gamma^2/\beta^2$, denotes the classical limit of the 
$d$-boson number operator $\hat{n}_d=\sum_\mu d^\dag_\mu d_\mu$ (restricted 
to $L=0$) and forms an isotropic harmonic oscillator Hamiltonian 
in the $\beta,\gamma$ variables. Notice that (\ref{eq:Hcl}) contains 
complicated momentum-dependent terms originating from the two-body 
interactions in~(\ref{eq:Hcri}), not just the usual 
quadratic kinetic energy $T$. 
Setting $p_{\beta} = p_\gamma=0$ in Eq.~(\ref{eq:Hcl}) leads to the 
potential 
\begin{eqnarray}
\label{eq:Vcl}
V_\mathrm{cri}/h_2 &=& 
{\textstyle\frac{1}{2}}\bz^2 \beta^2  
+ \textstyle{\frac{1}{4}}(1-\bz^2)\beta^4 \nonumber \\
&&- \textstyle{\frac{1}{2}}\bz \sqrt{2-\beta^2} \beta^3\cos{3\gamma} ~,
\end{eqnarray}
which can be alternatively obtained for $\gamma=0$ as an expectation value 
of $\hat{H}_\mathrm{cri}$~(\ref{eq:Hcri}) in the static condensate 
mentioned above, $V_\mathrm{cri}(\beta,\gamma=0)=
\bra{\beta;N}\hat{H}_\mathrm{cri}\ket{\beta;N}$.
Apart from being compact, the $\beta,\gamma$ variables are analogous to 
the coordinates of the geometric collective model~\cite{GCM} and portray 
the amount of quadrupole deformation ($\beta=0$ corresponds to the 
spherical, while $\beta>0$ to deformed shapes) and the triaxiality of the 
nucleus ($\gamma=2\pi k/3$, with $k=0,1,2$ correspond to prolate and 
$\gamma=\pi/3 + 2\pi k/3$ to oblate axially-symmetric shapes, while other 
values to triaxial shapes). They may be interpreted as polar coordinates 
in an abstract plane parametrized by Cartesian coordinates 
$x=\beta\cos{\gamma}$ and $y=\beta\sin{\gamma}$. The potential of 
Eq.~(\ref{eq:Vcl}) displays a three-fold rotational symmetry 
about the origin $\beta=0$. Its four minima, being degenerate at energy 
$V_\mathrm{min}=0$, are of two types and correspond to the different 
phases: (i)~the single spherical minimum at $\beta=0$ and (ii)~the 
three equivalent deformed minima at $\beta=\sqrt{2}\bz(1+\bz^2)^{-1/2}$ 
with $\gamma=0,2\pi/3,4\pi/3$. In between the spherical and deformed minima, 
there are three equivalent saddle points located at 
$\beta=[1-(1+\bz^2)^{-1/2}]^{1/2}$, which create 
potential barriers of height $V_b=h_{2}[1-(1+\beta_0^2)^{1/2}]^2/4$, 
separating the two phases. The limiting value at 
the domain boundary is $V(\beta=\sqrt{2},\gamma)=h_2$.  
The potential of Eq.~(\ref{eq:Vcl}) is similar to that considered 
in~\cite{ref:Bere04}, but differs by the square-root term and the 
compact domain.
 
The nature of the classical motion associated with the 
Hamiltonian~(\ref{eq:Hcl}) can be depicted 
conveniently via Poincar\'e surfaces of sections in the plane $y=0$, 
plotting the values of $x$ and the momentum $p_x$ each time a 
trajectory intersects the plane. 
Regular trajectories are bound to toroidal manifolds within the phase 
space and their intersections with the plane of section lie on 
1D curves (ovals). In contrast, chaotic trajectories randomly cover 
kinematically accessible areas of the section~\cite{ref:Reic92}. 

\begin{figure}[!ht]
\begin{center}
 \epsfig{file=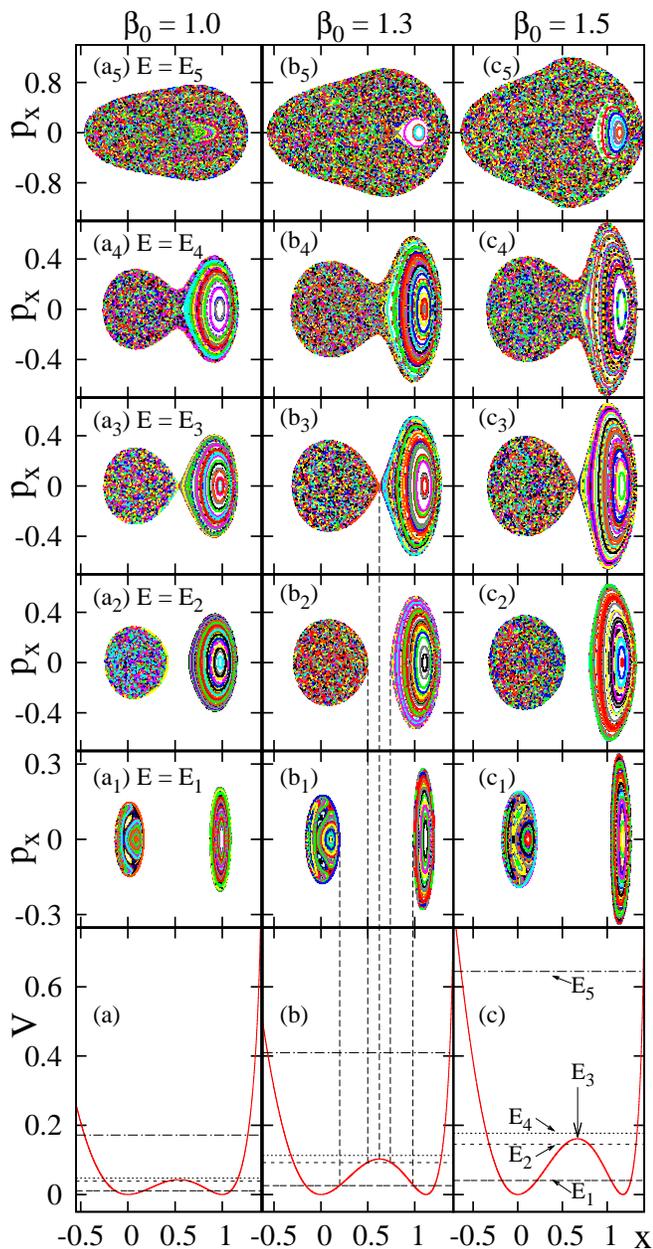,width=0.99\linewidth}
\end{center}
\caption{(Color online). 
Poincar\'e sections for the classical Hamiltonian~(\ref{eq:Hcl}) with 
$h_2=1$ and $\beta_0 = 1.0, 1.3, 1.5$ (left, middle and 
right columns, respectively). In each case, the surfaces are drawn 
at energies $E_k$ ($k=$ 1-5)
relative to the barrier-tops $V_b$: 
$E_1 = 0.25 {V}_b$, $E_2 = 0.9 {V}_b$, $E_3 = {V}_b$, $E_4 = 1.1 {V}_b$, 
$E_5 = 4 {V}_b$ (panels $a_k$-$b_k$-$c_k$).
The bottom row (panels $a$-$b$-$c$) depicts the corresponding classical 
potential $V_\mathrm{cri}(x,y=0)$, Eq.~(\ref{eq:Vcl}), with 
coexisting spherical and deformed minima at $x=0$ and $x\geq 1$, 
respectively. 
The reference energies, $E_k$, are indicated by horizontal dashed and dotted 
lines. Vertical lines mark the turning-points for 
$E_k\leq V_b$.
Notice the pronounced separation of regular and chaotic regions in phase 
space, associated with the two minima.} 
\label{fig:Fig1} 
\end{figure}

Poincar\'e sections for $\bz=1.0,1.3,1.5$, 
are displayed, respectively, in the left, middle and right 
columns of Fig.~\ref{fig:Fig1}. 
The three cases correspond to 
``low'', ``medium'' and ``high'' potential barriers 
$V_b/h_2=$ 0.04, 0.10, 0.16, 
(compared to $V_b/h_2=$ 0.0009 in previous works~\cite{ref:AW93}). 
In each case (see panels $a_k$-$b_k$-$c_k$), 
we show the sections at five different energies $E_k$ ($k=$1-5) 
relative to the barrier-tops $V_b$: 
$E_1 = 0.25 V_b$, $E_2 = 0.9 V_b$ are below, $E_3=V_b$ at, and 
$E_4=1.1 V_b$, $E_5=4 V_b$ are above the barrier. The bottom row 
(panels $a$-$b$-$c$) depicts the corresponding classical potential 
$V_\mathrm{cri}(x,y=0)$, Eq.~(\ref{eq:Vcl}). In all three cases, 
the motion is predominantly 
regular at low energies and gradually turning chaotic 
as the energy increases. However, the classical dynamics evolves 
differently in the vicinity of the two wells. 
The family of regular trajectories 
near the deformed minimum has a particularly simple 
structure. It forms a single set of concentric loops around 
a single stable (elliptic) fixed point. The trajectories remain regular even 
at energies far exceeding the barrier height $V_b$. In contrast, 
a more complicated structure develops near the spherical minimum. 
Here, at low energies (panels $a_1$-$b_1$-$c_1$), one observes
four major islands surrounding stable fixed points, and unstable 
(hyperbolic) fixed points in-between. 
As the energy increases (approximately for $E>V_b/3$~\cite{ref:Mace11}), 
considerable fraction of the trajectories becomes chaotic 
(panels $a_2$-$b_2$-$c_2)$ 
until complete chaoticity is reached near the barrier top. 
The clear separation between regular and chaotic dynamics, 
associated with the two minima, persists all the way to $E=V_b$, 
where the two regions just touch (panels $a_3$-$b_3$-$c_3$). 
At $E > V_b$, the chaotic trajectories from the spherical region can 
penetrate into the deformed region 
and a layer of chaos develops (panels $a_4$-$b_4$-$c_4$), 
and gradually dominates the surviving regular island 
for $E\gg V_b$ (panels $a_5$-$b_5$-$c_5$). In general, 
the regularity is more pronounced for higher barriers (larger $\bz$). 
\begin{figure}[!ht]
\begin{center}
\epsfig{file=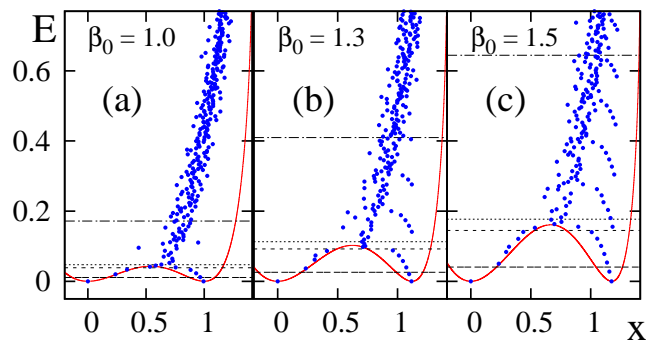,width=0.98\linewidth}
\end{center}
\caption{(Color online). 
Quantum Peres lattices $\{x_i,E_i\}$ for $L=0$ eigenstates $\ket{i}$
of $\hat{H}_\mathrm{cri}$~(\ref{eq:Hcri}), with $h_2=1$, $N=80$ and  
$\bz=1.0,1.3,1.5$.
The quantity $x_i= \sqrt{2\bra{i} \hat{n}_d\ket{i}/N}$ enables a 
direct comparison with the classical potential 
$V_\mathrm{cri}(x,y=0)$, Eq.~(\ref{eq:Vcl}). 
The reference energies $E_1,E_2,E_4,E_5$, are indicated 
as in Fig.~\ref{fig:Fig1}. 
Notice the sequences of regular states in the deformed region. 
The lowest sequence consists of 
$L=0$ bandhead states of 
the ground $g(K=0)$ and $\beta^n(K=0)$ bands. 
Regular sequences at higher energy correspond to 
$\beta^n\gamma^2(K=0)$, $\beta^n\gamma^{4}(K=0)$ bands, etc.} 
\label{fig:Fig2}
\end{figure}

Quantum manifestations of classical chaos are 
often detected by statistical analyses of energy 
spectra~\cite{ref:Reic92}. 
In a quantum system with mixed regular and irregular states, the 
statistical properties of the spectrum are usually intermediate between 
the Poisson and the Gaussian orthogonal ensemble (GOE) statistics. 
Such global measures of quantum chaos are, however, insufficient to 
reflect the rich dynamics of an inhomogeneous phase space structure, 
as in Fig.~1, with well separated regular and chaotic regions. 
To do so, one needs to distinguish between regular and irregular 
subsets of eigenstates in the same energy intervals. 
For that purpose we employ the spectral lattice method of 
Peres~\cite{ref:Pere84l}, which 
provides additional properties of individual energy eigenstates. 
The Peres lattices are constructed by plotting the expectation 
values $O_i = \bra{i}\hat{O}\ket{i}$ of an arbitrary operator, 
$[\hat{O},\hat{H}]\neq 0$, versus the energy $E_i=\bra{i}\hat{H}\ket{i}$ 
of the eigenstates $\ket{i}$. The lattices $\{O_i,E_i\}$ corresponding to 
regular dynamics can be shown to display a regular pattern, while chaotic 
dynamics leads to disordered meshes of points. 
The method has been recently applied to quantum 
chaos in the geometric collective model of nuclei~\cite{ref:Stran09}. 

In the present analysis we choose the Peres operator to be 
$\hat{O}=\hat{n}_d$. The lattices correspond to the set of~points 
$\{x_i,E_i\}$, 
with $x_i \equiv \sqrt{2 \bra{i}\hat{n}_d\ket{i}/N}$ and $\ket{i}$ being 
the eigenstates of $\hat{H}_\mathrm{cri}$~(\ref{eq:Hcri}). 
The expectation value of $\hat{n}_d$ in the condensate, 
${x} = \beta = \sqrt{2\bra{\beta;N}\hat{n}_d\ket{\beta;N}/N}$, 
is related to the deformation $\beta$ (whose equilibrium value 
is the order parameter of the QPT) and the coordinate 
$x$ in the classical potential, 
$V_\mathrm{cri}(x,y=0)=V_\mathrm{cri}(\beta,\gamma=0)$, Eq.~(\ref{eq:Vcl}). 
The spherical ground state has $n_d=x_i=0$. Excited spherical 
states show characteristic dominance of single $n_d$ 
components~\cite{ref:Lev06}, 
hence $x_i\sim \sqrt{n_d/N}$ is small for $n_d/N<<1$. 
Rotational members of the deformed ground band are obtained by 
$L$-projection from  $\ket{\beta;N}$ and have $x_i\approx \beta$ 
to leading order in $N$. 
This relation is still valid, to a good approximation, for states in 
excited deformed bands, whose intrinsic states are obtained 
by replacing condensate bosons in $\ket{\beta;N}$ with 
orthogonal bosons representing $\beta$ and $\gamma$ 
excitations~\cite{lev87}. 
These attributes have the virtue that the lattices $\{x_i,E_i\}$ 
can identify the regular/irregular quantum states and 
associate them with a given region in the classical phase space.

Fig.~\ref{fig:Fig2} presents the lattices calculated for $L=0$ 
eigenstates of $\hat{H}_\mathrm{cri}$ with $N=80$ and $\bz=1.0,1.3,1.5$. 
In each case, one can clearly identify regular sequences of 
states localized within and above the respective 
deformed wells. They form several chains of lattice points 
close in energy, with the lowest chain 
originating at the deformed ground state. 
A close inspection reveals that the $x_i$-values 
of these regular states, lie in the intervals of $x$-values 
occupied by the regular tori in the Poincar\'e sections 
in Fig.~\ref{fig:Fig1}. Similarly to the classical tori, these regular 
sequences persist to energies well above the barriers $V_b$. 
In contrast, the remaining states, 
including those residing in the spherical minimum, 
do not show any obvious patterns and lead to 
disordered (chaotic) meshes of points at high energy $E>V_b$. 
(Although there are only a few lattice points 
near $x=0$, the chaotic behavior of these spherical 
states is evident by glancing at the Poincar\'e sections 
of Fig.~\ref{fig:Fig1}). 
The ability of the Peres method to separate regular states from irregular 
ones can be tested by fitting the nearest neighbors level 
spacing distribution of the quantum spectrum by a Brody 
distribution $P_\omega(S)$~\cite{ref:Reic92}. 
The parameter $\omega$ interpolates between 
the Poisson ($\omega=0$) and GOE statistics ($\omega=1$), 
corresponding to integrable and fully chaotic classical motion, 
respectively. We determine $\omega$ for the full (mixed) 
spectrum of $L=0$ eigenstates and for a partial (``chaotic'') spectrum, 
from which the tentatively regular levels are excluded. 
Specifically, for $(N=80,\beta_0=1.5)$, we eliminate the $32$ levels which 
form the regular sublattice in the deformed phase (see Fig.~\ref{fig:Fig2}c) 
from a set of $364$ states with $E<0.8$. For the mixed spectrum we obtain 
$\omega = 0.78 \pm 0.10$ and for the ``chaotic'' spectrum 
$\omega = 1.01 \pm 0.10$, thus confirming the regular character of the 
excluded levels. 
\begin{figure}[t]
\begin{center}
\epsfig{file=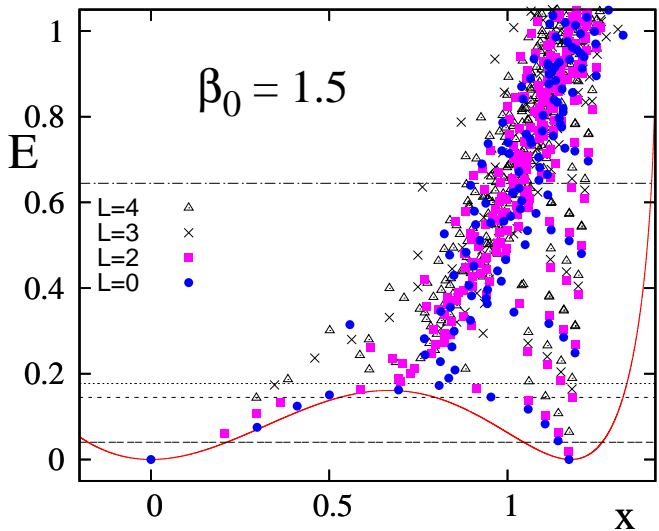,width=1\linewidth}
\end{center}
\caption{(Color online). Quantum Peres lattices $\{x_i,E_i\}$ 
for eigenstates of $\hat{H}_\mathrm{cri}$~(\ref{eq:Hcri})
with $h_2=1,\,N=50,\,\beta_0=1.5$ and 
$L=0,2,3,4$. To enhance visibility, a small $\hat{L}^2$ term is added, 
leading to an energy shift $\Delta E_L = 0.0032 L(L+1)$, 
without affecting the wave functions. The classical potential 
$V_\mathrm{cri}(x,y=0)$ and reference energies $E_k$ are indicated 
as in Fig.~\ref{fig:Fig2}. Notice the rotational bands 
($K=0$, $L=0,2,4$) and ($K=2$, $L=2,3,4$)
formed by the regular states in the deformed phase.}
\label{fig:Fig3}
\end{figure}

The Peres lattices 
corresponding to eigenstates of 
$\hat{H}_\mathrm{cri}$~(\ref{eq:Hcri}) with 
$L=0,2,3,4$, $N=50$ and $\beta_0=1.5$ 
are shown in Fig.~\ref{fig:Fig3}. 
They disclose families of regular patterns in the deformed phase. 
These correspond to rotational bands of states having a common intrinsic 
structure, as indicated by their nearly equal values of 
$\langle \hat{n}_d \rangle$. The regular $L=0$ states form 
bandheads of rotational sequences $L=0,2,4,\ldots$ ($K=0$ bands) and 
are accompanied by sequences $L=2,3,4,\ldots$ built upon regular $L=2$ states 
($K=2$ bands). Additional $K$-bands with $L=K,K+1,K+2,\ldots$ (not shown in 
Fig.~\ref{fig:Fig3}) can also be identified. In the nuclear physics 
terminology, the lowest $K=0$ band refers to the ground band and 
excited $K$-bands correspond to multiple $\beta$ and $\gamma$ vibrations 
about the deformed shape with angular momentum 
projection $K$ along the symmetry axis.
Detailed analysis reveals additional 
band-like properties of these states, in particular, the coherent 
decomposition of wave functions in the rotor basis~\cite{ref:Mace11}.  
Such ordered band structures are not present in the chaotic parts 
of the lattice 
and their persistence in the spectrum is unexpected, in view of 
the strong mixing and abrupt structural changes taking place at the 
critical point. It validates 
the recently proposed regularity-induced adiabatic separation of 
intrinsic and collective dynamics~\cite{ref:Mace10}, 
for a subset of states.

In summary, we have examined the interplay of regularity and chaos in 
an interacting system modeling the dynamics at the critical point 
of a first order QPT between spherical and deformed nuclei. 
A classical analysis, using Poincar\'e sections, shows that 
the two coexisting phases exhibit different 
susceptibilities towards the onset of chaos. 
While the deformed phase displays robustly ordered and 
rather simple dynamics, 
the spherical phase shows strongly chaotic behavior. 
An analysis of the quantum spectrum, using Peres lattices, 
identifies unexpected regular rotational bands in the deformed region, 
which are absent from the disordered portions of the lattice. 
Although the regular and chaotic motions coexist in a broad interval of 
energies, they are well separated and can be distinguished by both 
classical and quantum calculations. 
The system in the domain of phase coexistence, 
thus provides a clear cut demonstration of the classical-quantum 
correspondence of regular and chaotic behavior, illustrating Percival's 
conjecture concerning the distinct properties of regular and irregular 
quantum spectra~\cite{ref:Perc73}. 

One of us (M.M.) wishes to thank to P. Cejnar for useful discussions. 
This work is supported by the Israel Science Foundation.

\end{document}